Title: High-spatial resolution, high-quality white matter tractography using DTI with dynamic slice-by-slice B0 shimming in a head-only high-gradient performance 3T MRI scanner.

Authors: Jerome J. Maller[1*], PhD, Patricia Lan[2], PhD, Sherry S. Huang[3], PhD, Chitresh Bhushan[4], PhD, Xinzeng Wang[5], PhD, Ante Zhu[4], PhD, Vincent A. Magnotta[1], PhD.

[1]Department of Radiology, Iowa University, Iowa City, Iowa, USA

[2]MR Clinical Solutions & Research Collaborations, GE HealthCare, Menlo Park, CA, USA

[3]Science and Technology Office, GE HealthCare, Rochester, MN, USA

[4]Technology & Innovation Center, GE HealthCare, Niskayuna, New York, USA

[5]MR Clinical Solutions & Research Collaborations, GE HealthCare, Houston, Texas, USA

*Corresponding author information: Jerome J Maller, Department of Radiology, University of Iowa, 169 Newton Road, Iowa City, IA 52242

E-mail: jerome-maller@uiowa.edu

## Abstract

Data from conventional diffusion tensor imaging (DTI) using single-shot echo planar imaging (SS-EPI) acquisition are substantially influenced by magnetic field inhomogeneities ($\Delta B_0$), which result in image distortion and signal dephasing. Multi-shot EPI can mitigate the $\Delta B_0$-induced artifacts but at the cost of increased scan time. Recent brain tissue-selective, dynamic slice-by-slice $B_0$ shimming techniques have demonstrated effective reduction of local $\Delta B_0$ in the brain. When combined with DTI utilizing an echo-planar imaging (EPI) read-out, image distortion and signal dephasing are significantly reduced. The purpose of our study was to assess the impact of DTI with dynamic slice-by-slice $B_0$ shimming technique (DySiBo) on white matter (WM) tractography. We retrospectively analyzed DTI datasets using SS-EPI and Multi-shot EPI with 2 shots (MS-2shot-EPI) readout at two spatial resolutions (1x1x2mm$^3$ and 2x2x2mm$^3$) with and without DySiBo. WM tractography was generated and inspected to examine the impact upon regions typically affected by $\Delta B_0$-induced artefacts. We found more WM streamlines were generated in the DTI datasets with DySiBo in regions conventionally impacted by $\Delta B_0$-induced in EPI images, including the brainstem, temporal and frontal lobes. DySiBo substantially improved the generation of WM streamlines in DTI-based tractography at both in-plane resolutions. In conclusion, DySiBo in conjunction with DTI requiring 10 diffusion tensor directions is sufficient to generate data with high enough SNR and angular resolution to resolve crossing fibers and produce high quality WM tractography in brain regions typically affected by susceptibility-induced artefacts. This has implications for quantitative WM microstructural indices and clinical evaluation of WM tracts in patients.

## 1 Introduction

Diffusion MRI (dMRI) colloquially known as diffusion tensor imaging (DTI; [1]) has been widely used to non-invasively probe white matter (WM) microstructure in vivo. In the research domain, it can offer insight into tissue microstructural changes which can be used to infer differences in connectivity when comparing to relatively large cohorts [2, 3]. In the clinic, dMRI can be used to investigate regions of reduced white matter microstructure which has been shown to assist in treatment planning [4]. Renderings of dMRI-based tractograms using three-dimensional streamlines offers additional information especially in the neurosurgical arena.

Conventional dMRI data are generated using a single-shot echo planar imaging (SS-EPI) technique which is subject to B0 field inhomogeneity ($\Delta B_0$)-induced susceptibility artefacts in regions of air/bone interfaces, such as the brainstem, temporal lobe, and frontal lobe [5]. Those susceptibility artefacts in EPI include image distortion, signal dephasing, and/or signal pile-up along the phase-encoding direction. Susceptibility related artefacts have been shown to lead to inaccurate WM fiber tracking [6]. Furthermore, these artefacts are exacerbated with increasing magnetic strength (e.g. comparing 1.5T to 7.0T). The susceptibility artefacts become more severe as spatial resolution increases because of the longer echo spacing which reduces the effective pixel bandwidth in the phase-encoding direction. Therefore, achieving high-spatial resolution dMRI with high image quality has been a common challenge in neuroimaging.

Software-based approaches have been proposed to mitigate $\Delta B_0$-induced artifacts in dMRI-based neuroimaging, which have been demonstrated as effective and robust, and made commercially available for clinical use. The most widely used technique is to acquire a volume with b-value of zero (i.e., $T_2$-weighted) in the opposite phase-encoding direction. This creates maps of frequencies specific to regions of $\Delta B_0$ field, which can be used to quantify the susceptibility artefacts and subsequently used to correct for image distortion [7]. This technique is referred to as 'blipping' and attempts to correct for $\Delta B_0$-induced artefacts retrospectively after the data acquisition. Another approach to minimize $\Delta B_0$-induced artefacts is to prospectively acquire images with multi-shot EPI (MS-EPI) acquisition [8]. However, multi-shot EPI requires longer acquisition time as well as increased likelihood of motion-induced artifacts.

In addition, other approaches that require additional hardware have been also applied to reduce $\Delta B_0$-induced artifacts in high-resolution dMRI-based neuroimaging. For example, multi-coil shimming and integrated RF and $B_0$ shim array have been developed, dedicated for neuroimaging [9-11]. Multi-channel $B_0$ shim coils (or integrated with RF coils) and drivers are used to generate high-order magnetic field shimming to reduce $\Delta B_0$, which is more efficient than the zero and first-order $B_0$ shimming using imaging gradients of X, Y, and Z coils and/or the high-order shimming coils integrated with the gradient coil. However, these approaches require additional and expensive hardware, and have only been demonstrated in a few research sites.

Dynamic slice-by-slice, zero and first-order $B_0$ shimming techniques have been shown to be an effective approach to prospectively minimize local $\Delta B_0$ and $\Delta B_0$-induced susceptibility artefacts in each 2D slice of whole-brain dMRI acquisitions [12-16]. In comparison, standard $B_0$ shimming for dMRI acquisitions is static and minimizes $\Delta B_0$ across the whole-brain, which is not effective in minimizing local $\Delta B_0$. Importantly, these techniques are compatible with any MRI systems without requiring expensive hardware such as high-order $B_0$ shimming coils and drivers, making it feasible for wide utilization within both research and clinical settings. These techniques can also work with any RF coils from single-channel RF transmit/receive coil to multi-channel RF coils. Furthermore, it is also compatible with single-shot and multi-shot EPI acquisitions, and retrospective correction methods such as ‘blipping. By combining all of these techniques, Lan et al have shown minimal image distortion and signal dephasing in spin-echo echo-planar images and the resulting diffusivity maps at 1 mm isotropic resolution for the brain [17].

Recent development of a brain tissue-selective, dynamic slice-by-slice, zero and first order $B_0$ shimming technique (named as “DySiBo” in this work) has been proposed to further reduce $\Delta B_0$ in brain tissue regions [17]. In comparison, standard $B_0$ shimming reduces $\Delta B_0$ across the whole head, which was shown to not effectively reduce $\Delta B_0$ in the primary region-of-interest (e.g., brain tissue only). Furthermore, DySiBo also accounted from gradient nonlinearity in the calculation of shimming coefficients, which has been shown to be important especially in off-center regions where the nonlinearity of imaging gradient coils cannot be ignored [17]. We have previously reported an initial demonstration of improved accuracy of diffusivity characterization (including fractional anisotropy as well as mean/axial/radial diffusivity) of the brain in vivo using a head-only high-performance gradient human 3T MRI system [18]. The high slew rate and high peripheral nerve stimulation threshold of a head-only high-performance gradient MRI system shortens echo spacing in EPI, resulting in less susceptibility artefacts compared to whole-body MRI systems. Adding DySiBo to the echo-planar imaging sequence has been shown to further reduce image distortion or signal dephasing and thus improve diffusivity measurements, especially at high-spatial resolution (e.g. 1x1 $mm^2$ in-plane). However, the impact of DySibo on dMRI-based tractography was not evaluated.

In this work, we retrospectively perform tractography on a high-resolution single shot EPI (SS-EPI) and multi-shot EPI (MS-EPI) dMRI data to investigate whether DySiBo improves tractography streamline generation in regions typically effected by $B_0$ field inhomogeneities.

# 2 Materials and Methods

## 2.1 Participants

Informed consent was obtained for a single subject (34 years old, female) who was scanned under a local institutional review board-approved protocol using an investigational MAGNUS head-only 3T MRI system equipped with a high-performance gradient system: maximum gradient amplitude of 300 mT/m and slew rate of 750 T/m/s [19]. A 32-channel receive radiofrequency coil (Nova Medical, USA) was used for the study.

## 2.2 MRI acquisition and processing

The imaging protocol consisted of a 3-plane localizer, 2D $B_0$ field mapping using a gradient-echo sequence, and T1-weighted MPRAGE scan collected at 1mm isotropic resolution. In addition, eight dMRI with 2mm slice thickness were acquired with two in-plane resolutions (1mm x 1mm, and 2mm x 2mm). The dMRI imaging data was collected at each resolution using SS-EPI and MS-EPI (2 shots) with and without DySiBo applied (Table 1). All dMRI data were acquired using pulsed gradient spin echo (PGSE) diffusion encoding at 10 diffusion directions and a b-value of 1000 s/mm$^2$. A single b=zero volume was also acquired.

**Table 1. about here**

Each of the dMRI datasets were reconstructed using a customized deep learning phase correction technique [20-23] integrated with a commercialized deep learning model for denoising and deringing [23].

## 2.3 White matter tract reconstruction

After image reconstruction, the resulting dMRI data was analyzed using an established pipeline as follows. Image processing began with correction for eddy current and movement distortions [24, 25] followed by DICOM to NIfTI format conversion using dcm2niix [26]. NIfTI files were then processed in MRtrix3 [27] by creating response files [28] and fiber orientation distributions (FODs) [29]. The T1-weighted MPRAGE image was then registered to each dMRI scan using FLIRT [30] from which WM maps were generated using the FAST module (Fully Automated Segmentation Tool; [31]). Those WM maps were then used as the seeds and masks to generate CSD (constrained spherical deconvolution; [32]) probabilistic tractography (250 thousand streamlines) based on iFOD2 [33] using default settings. Each tractography file was also visualized using photo-realistic rendering (Vibrant Tractography; [34]) to allow for higher contrast inspection of the streamlines.

## 2.4 $B_0$ field inhomogeneity and voxel displacement measurements

$B_0$ field inhomogeneity in the unit of frequencies (Hz), as well as voxel displacements in echo planar images, were calculated for field maps resulting from the acquisitions collected with and without DySiBo using imaging parameters from EPI acquisitions as follows:

Voxel displacement = $\Delta B_0$ x Echo-spacing x $N_{PE}$ / number-of-shots,

where $N_{PE}$ is the number of phase encoding lines for the full k-space of EPI.

# 3 Results

## 3.1 Tractography

Tractography data generated from the imaging sequences employing DySiBo resulted in a similar number of streamlines across the whole brain when compared to DySiBo off, with a maximum difference of less than 1% across the acquisitions. However, the acquisitions employing DySiBo resulted in more fibers in regions known to have a large magnetic susceptibility variation: inferior medial temporal lobe, anterior cerebellar peduncle, brainstem, and orbitofrontal regions (Figure 1). Using the cerebellar peduncle as a structure to quantify the changes in the number of fiber tracts, the acquisitions utilizing DySiBo shimming resulted in 7.3% to 26.4% more fiber tracts within this region (Table 2). This was evident at both 1mmx1mm (Figures 1A and 1C) and 2mmx2mm (Figures 1B and 1D) in-plane resolutions. However, the improvement was not uniform across resolutions and shots. That is the single shot data had the largest increase in the number of streamlines for the 1mm in-plane resolution while at 2mm in-plane resolution, the multi-shot data had the largest increase in the number of streamlines when DySiBo was employed.

**Table 2. about here**

Similar quality improvements were also observed in the streamlines located in the frontal lobe. Figure 2 shows that some frontal lobe streamlines are represented only in the data collected using DySiBo. Figure 2 shows a comparison between the two 1mm in-plane acquisitions using DySiBo (Fig 2 left) that were not present using the SS-EPI data collected without DySiBo (Figure 2, right). The same observations also existed in the 2mm in-plane data but is not shown here.

DySiBo employed for the 1mm in-plane SS-EPI acquisition (Figure 1A) did not eliminate all susceptibility related artefacts but it did minimize their effects allowing for many more streamlines to be generated in white matter regions near regions with large susceptibility variations. Furthermore, the resulting dMRI data collected using DySiBo better aligned with the corresponding T1-weighted anatomical image (Figure 1).

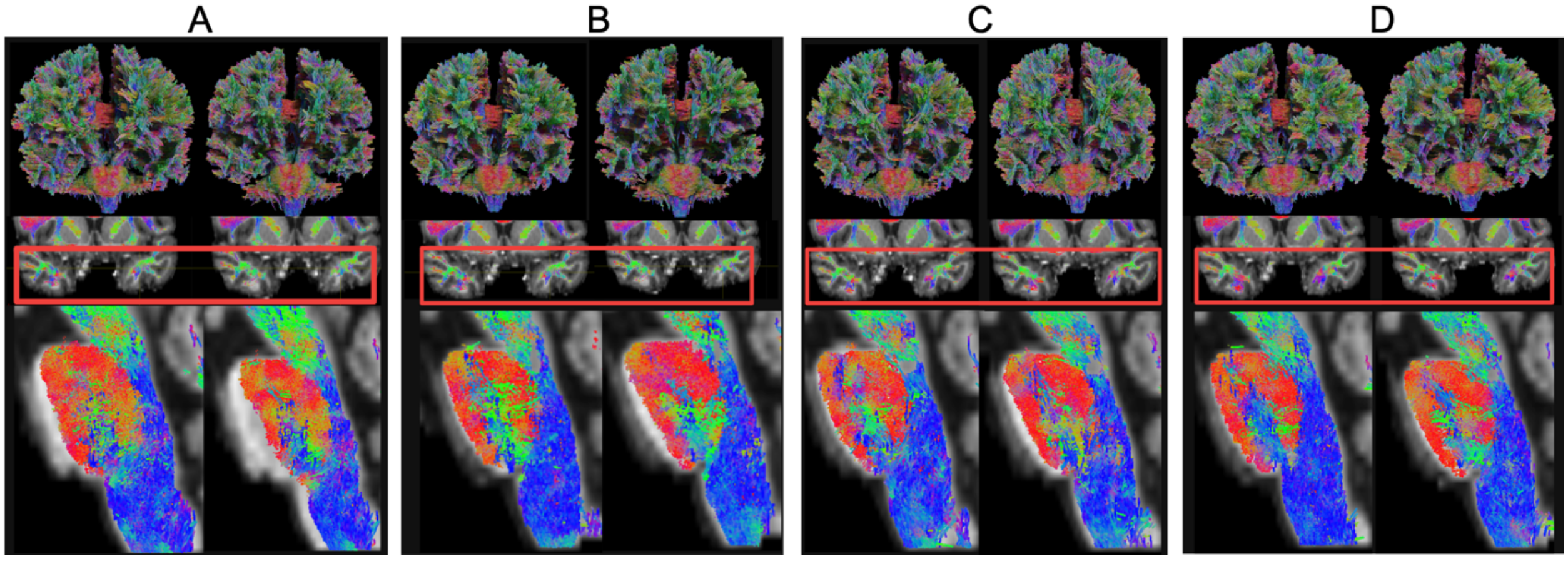

Figure 1. A. 1mmx1mm SS-EPI tractography (Left=DySiBo, Right=no DySiBo). B. 2mmx2mm SS-EPI tractography. C. 1mmx1mm MS-2shot-EPI tractography. D. 2mmx2mm MS-2shot-EPI tractography. The background gray-scale images were T1-MPRAGE. Top=Three-dimensional whole-brain rendering. Middle=Anterior temporal lobe from coronal. Bottom=Brainstem from sagittal.

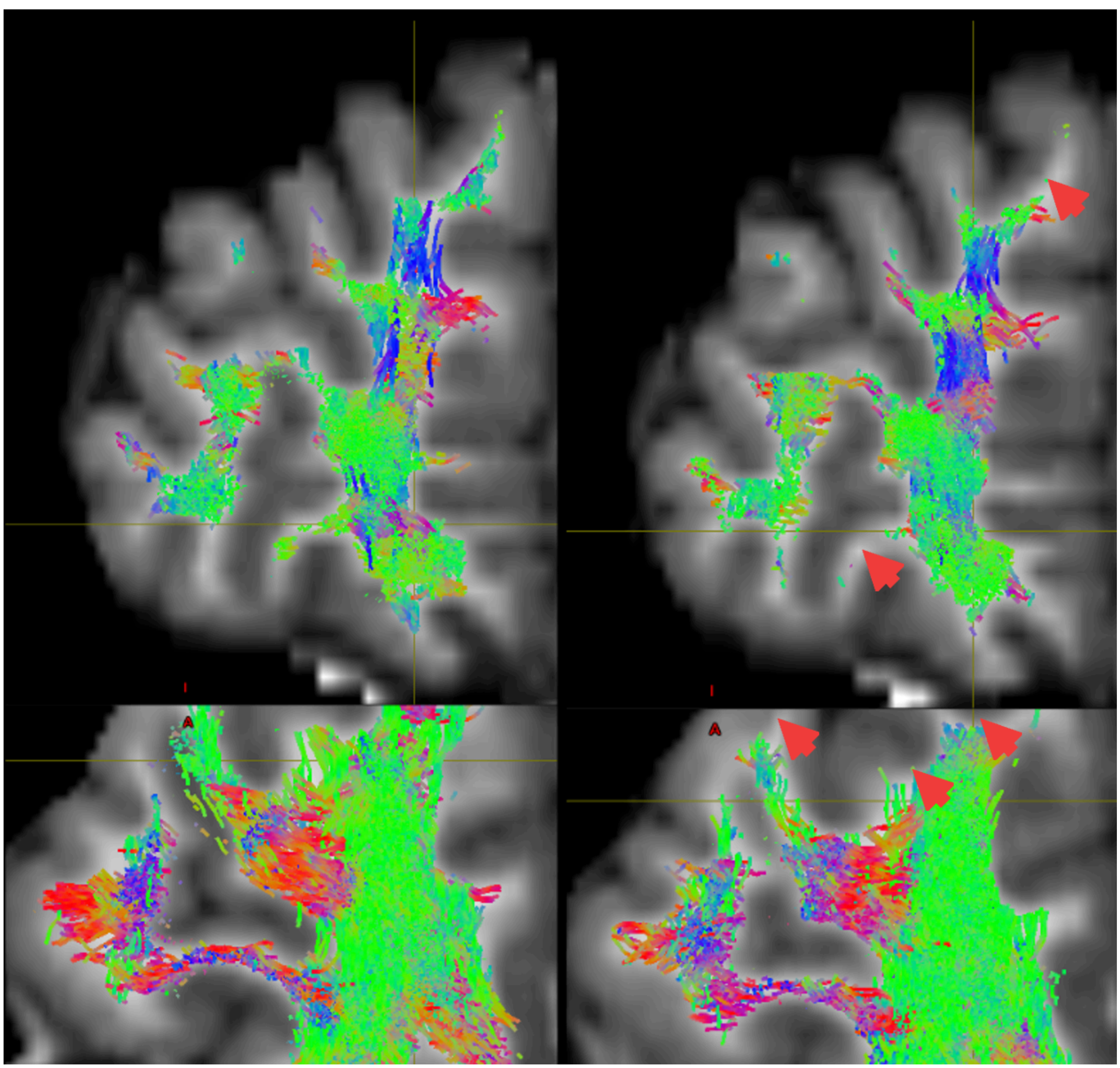

Figure 2. Frontal lobe streamlines from SS-EPI 1mmx1mmx2mm data. Left: using DySiBo; Right: without using DySiBo. The red arrows indicate streamlines that are missing in the data collected without DySiBo. Top=Coronal view, Bottom=Axial view.

### 3.2 $B_0$ field inhomogeneity and voxel displacement

The resulting improvement in the ability to perform streamline tractography with the acquired data appears to result from reduced variation $B_0$ variation within the brain tissue. A dual echo $B_0$ field mapping sequence was acquired with and without DySiBo shimming for the same slices acquired as part of the diffusion imaging experiment (Table 3). $B_0$ field inhomogeneity, as measured from $B_0$ field maps, was reduced by DySiBo, with both mean (17% improvement) and median (88% improvement) values showing decreases when DySiBo was applied. The mean voxel displacement was reduced by 17.5% and the median by 50% when using DySiBo shimming (Table 4).

**Table 3. about here**

**Table 4. about here**

Finally, we investigated post-hoc the ability of data to resolve crossing fibers in the region where the corpus callosum, corticospinal tract, and arcuate fasciculus are present within the same voxels. The FODs for the same voxel is shown for the 1mm and 2mm isotropic data in Figure 3 for the data collected using DySiBo shimming. Sharp FODs were clearly visible in both datasets contain these crossing fibers (Figure 3). This was possible due to the high SNR data that was available at 2mm$^3$ voxel volume.

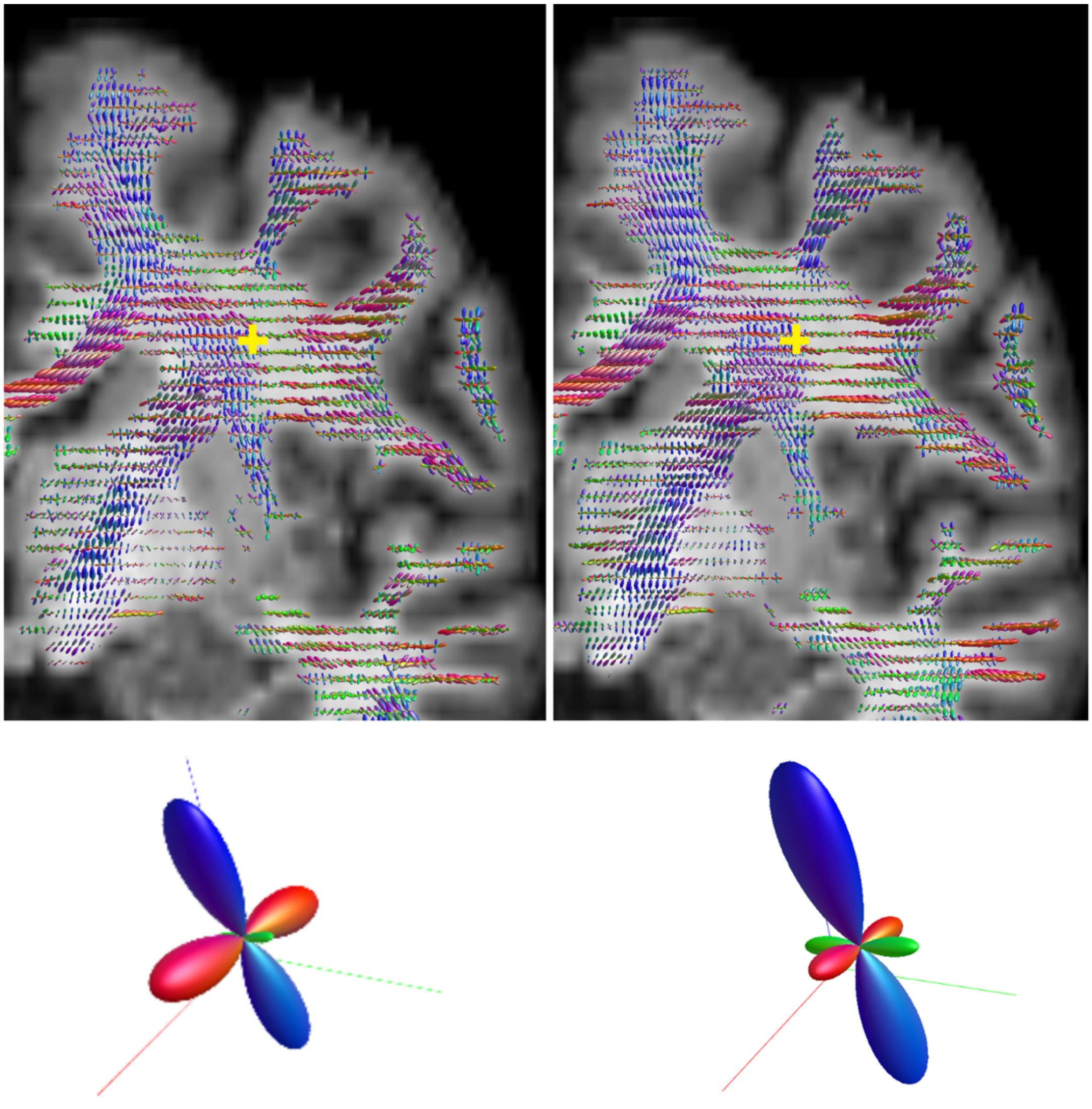

Figure 3. Resolved fiber orientation distributions (FODs) in a voxel comprised of crossing fibers from the tapetum of the corpus callosum (red), corticospinal tract (blue), arcuate fasciculus (green). Left: SS-EPI (1mmx1mmx2mm), Right: 2 shot MS-EPI (2mmx2mmx2mm).

## 4 Discussion

Our study compared dMRI datasets using SS-EPI and MS-EPI (2 shots) readout at two in-plane spatial resolutions (1mm and 2mm) with and without brain tissue-selective, dynamic slice-by-slice $B_0$ shimming technique (DySiBo). The results demonstrated that DySiBo results in less distortion leading to more streamlines propagating into regions typically affected by $B_0$ field inhomogeneities.

MS-EPI (e.g. MUltiplexed Sensitivity-Encoding; MUSE [8]) helps reduce susceptibility artefacts and provides significantly greater SNR than conventional SS-EPI [35]. By utilizing segmented readout, susceptibility-induced artefacts are greatly reduced in brain regions such as the brainstem, temporal and frontal lobes. Furthermore, as the number of shots increases the susceptibility related artefacts decrease. In the current study, we used a 2-shot sequence, which we found to be sufficient to prevent nearly all signs of susceptibility artefact in the high-performance gradient head-only MRI system where the echo spacing is almost half of that in clinical whole-body MRI system. This was supported by the findings of reduced voxel displacement by using DySiBo (17% mean and 48% median reduction). The use of DySiBo coupled with MS-EPI slightly improved the results, especially at relatively low spatial resolution of 2mm in-plane, since very few artefacts existed in this acquisition even without applying DySiBo.

The results of this retrospective study demonstrated that high-quality tractography can be confidently achieved using SS-EPI acquisition with DySiBo without suffering from motion-induced artifacts that can occur using the MS-EPI approach. Furthermore, “blipping”-based distortion correction can be leveraged to improve image quality in combination with DySiBo, as previously demonstrated [17]. The high resolution tractography was possible due to the high quality data collected using the DySiBo approach. The resulting tractography information was able to generate WM tracks even in the brainstem where susceptibility related artifacts have made such tract generation challenging. This improvement in the ability to generate high quality tractography data will help improve our understanding of human brain connectivity and its alternation in clinical populations.

The landmark article by Derek Jones [36] concluded from a Monte Carlo simulation that at least 30 unique gradient directions are required for a robust estimation of tensor-orientation and at least 20 unique sampling orientations for robust anisotropy measurements at 3T. Kumpulainen et al. (2022; [37]) reported a minimum of 18 directions are required to obtain reliable scalar values at 3T, while Lebel et al. (2012; [38]) showed that six directions provides comparable robustness to 30- or 60-direction data for deterministic tractography and anisotropy measures. Hence, our dMRI scheme using only 10 diffusion directions is sufficient to generate reliable scalar values in DTI at b of 1000 s/mm$^2$ acquired on a high-performance gradient head-only MRI system at 3T, which operates at maximum gradient strength of 300 mT/m and maximum slew rate of 750 T/m/s.

In whole-body 3T MRI systems, it is generally accepted that at least 30 unique diffusion directions are required to generate accurate tractography, and at least 60 to resolve crossing and kissing/abutting fibers, although some have concluded that 20 may suffice [39-41] using advanced models that can estimate multiple fiber orientations within a single voxel, such as probabilistic tractography based on convoluted spherical deconvolution [29]. Notably, previous studies estimating the minimum number of directions required to resolve crossing fibers were in the context of SNR available from conventional 3T whole-body MRI scanners with gradients with a maximum amplitude on the order of 50 mT/m and slew rate of 200 T/m/s. The increase in TE (which at b=1000 s/mm$^2$ is typically 70-90ms in a conventional whole-body 3T and 54-60ms in the MAGNUS) due to gradient performance on such scanners results in significant reductions in SNR making it difficult to resolve small crossing-angles [42].

With the advent of head-only MRI scanners with high-performance gradients (in our case, 300 mT/m and 750 T/m/s) offering substantially shorter echo times (approximately half) and reduced echo spacing, the SNR at even higher spatial resolution substantially increases therefore resulting in a reduction of the minimum required diffusion directions needed to resolve crossing fibers. As such, it is not surprising that we were able to demonstrate crossing fibers in our data which comprised only 10 unique gradients diffusion directions, and not unexpected that the data even with the lowest SNR (SS-EPI 1mmx1mm) had similar amplitude FODs when compared to the MS-EPI 2mmx2mm data. DySiBo did not impact the FODs in these regions likely due to the fact that they are well away from regions with large susceptibility related artifacts.

Maximizing SNR and minimizing scan time were not the primary aims of the study. Rather, we aimed to investigate whether DySiBo would improve tractography accuracy in single-shot and multi-shot acquisitions and at two different spatial resolutions. Our future direction will focus on reducing TE to increase SNR and use acceleration techniques to decrease scan times.

There are certain limitations of the study. Firstly, we scanned only a single subject hence the generalizability of the study is limited although we are currently working on further improvement of the technique which we will apply to a group of participants in the future. We also did not acquire reverse phase-encoded B0 data thus we could not compare DySiBo to standard techniques to correct for susceptibility-induced distortion such as topup [7].

## 5 Conclusions

We demonstrated that using DySiBo in conjunction with DTI on a high-gradient performance 3T MRI system requiring 10 diffusion tensor directions in less than a 4 minute acquisition is sufficient to generate data with high enough SNR and angular resolution to resolve crossing fibers and produce high quality WM tractography in brain regions typically affected by susceptibility-induced artefacts. This is apparent for both typical $8mm^3$ and higher resolution $2mm^3$ voxel volumes. DySiBo delivers technical and translational value for high-spatial resolution, high-quality imaging of white matter tracks. For research, this offers the opportunity to reduce scanning times, providing more robust image quality especially in subjects who may have issue staying still during long scans. In the clinical arena, this translates to repeatable image quality and reliable diagnosis and/or treatment planning for patients who benefit from tractography evaluation. Moreover, the benefit of scanning time reduction is vital for adherence in patients with claustrophobia. In a hospital emergency department environment where MRI scanning time is at a premium, this sequence coupled with high performance gradients could provide substantial improvements to patient management.

Funding: This study was funded by NIH S10 Grant number S10OD030220 and U01EB034313.

## Tables

Table 1. MR Imaging Parameters.

| | $B_0$ field mapping | DTI | |
|---|---|---|---|
| **Shots** | N/A | 1 / 2 | 1 / 2 |
| **Spatial resolution** | 3.4 x 3.4 x 2mm$^3$ | 1 x 1 x 2mm$^3$ | 2 x 2 x 2mm$^3$ |
| **In-plane acceleration factor** | 1 | 2 | 2 |
| **Echo spacing** | N/A | 576 / 288 µs | 344 / 172 µs |
| **TE** | 2 and 3.7 ms | 61.8 / 62.1 ms | 53.6 / 54.1 ms |
| **TR** | 300 ms | 7700 / 7200 ms | 6800 / 6500 ms |
| **Flip angle** | 10$^o$ | 90$^o$-180$^o$ | 90$^o$-180$^o$ |
| **b-value** | N/A | 1000s/mm$^2$ | 1000s/mm$^2$ |
| **Number of averages** | N/A | 1 | 1 |
| **# of encoding directions** | N/A | 10 | 10 |
| **Scan time** | 56 sec | 1:31 / 2:53 mins | 1:22 / 2:36 mins |

Table 2. Number of whole brain and middle cerebellar peduncle streamlines reconstructed from DTI with DySiBo on and off for each of the acquisitions with different spatial resolutions and number of shots.

| | DySiBo ON | | DySiBo OFF | | Difference (%) | |
|---|---|---|---|---|---|---|
| | Whole-brain | MCP | Whole-brain | MCP | Whole-brain | MCP |
| **1mm (SS)** | 307845 | 11232 | 308581 | 8613 | -736 (-0.02) | 2619 (26.4) |
| **2mm (SS)** | 294527 | 11311 | 297280 | 10453 | -2753 (-0.09) | 858 (8.9) |
| **1mm (MS)** | 307758 | 9519 | 307574 | 8849 | 184 (0.01) | 670 (7.3) |
| **2mm (MS)** | 294531 | 11476 | 296902 | 9334 | -2371 (-0.01) | 2152 (20.6) |

MCP=Middle cerebellar peduncle; MS=Multi-shot; SD=Standard deviation; SS=Single-shot.

Table 3. $B_0$ field inhomogeneity from $B_0$ field maps for dynamic shimming on and off.

| | **DySiBo ON** | **DySiBo OFF** | **Difference (%)** |
|---|---|---|---|
| **Hz Mean (SD)** | 8.48 (28.16) | 10.13 (45.57) | -1.65 (-17.7) |
| **Hz Median** | 11 | 18 | -7 (88) |

Hz=Hertz; SD=Standard deviation.

Table 4. Voxel displacements for dynamic shimming on and off for each of the spatial resolutions and shot-factor acquisitions.

| | **DySiBo ON** | | | | **DySiBo OFF** | | | | **Difference** | | | |
|---|---|---|---|---|---|---|---|---|---|---|---|---|
| | VD Mean (SD) | | VD Median | | VD Mean (SD) | | VD Median | | Mean (%) | | Median (%) | |
| | Voxels | mm | Voxels | mm | Voxels | mm | Voxels | mm | Voxels | mm | Voxels | mm |
| **1mm (SS)** | 1.25 (4.14) | 1.25 (4.14) | 1.62 | 1.62 | 1.49 (4.30) | 1.49 (4.30) | 2.65 | 2.65 | -0.24 (-17.5) | -0.24 (-17.5) | -1.03 (-48.2) | -1.03 (-48.2) |
| **2mm (SS)** | 0.74 (2.44) | 1.48 (4.88) | 0.96 | 1.92 | 0.88 (2.54) | 1.76 (5.08) | 1.57 | 3.14 | -0.14 (-17.3) | -0.28 (-17.3) | -0.61 (-48.2) | -1.22 (-48.2) |
| **1mm (MS)** | 0.62 (2.07) | 0.62 (2.07) | 0.81 | 0.81 | 0.74 (2.15) | 0.74 (2.15) | 1.33 | 1.33 | -0.12 (-17.6) | -0.12 (-17.6) | -0.52 (-48.6) | -0.52 (-48.6) |
| **2mm (MS)** | 0.37 (1.22) | 0.74 (2.44) | 0.48 | 0.96 | 0.44 (1.27) | 0.88 (1.54) | 0.78 | 1.56 | -0.07 (-17.3) | -0.14 (-17.6) | -0.30 (-47.6) | -0.60 (-47.6) |

mm=millimeters; MS=Multi-shot; SD=Standard deviation; SS=Single-shot; VD=Voxel displacement.